\RequirePackage{fix-cm}
\documentclass[smallextended]{svjour3}       

\smartqed  
\usepackage{graphicx}
\usepackage{rotating}
\usepackage[usenames]{xcolor}

\usepackage{hyperref}
\usepackage{changes}
\definechangesauthor[name=John L,color=red]{JL}
\definechangesauthor[name=Vincenzo A,color=green]{VA}

\journalname{Experimental Astronomy}
\begin{document}

\title{A journey of exploration to the polar regions of a star: probing the solar poles and the heliosphere from high helio-latitude }

\titlerunning{A journey of exploration to the polar regions of a star}        

\author{Louise~Harra \and
Vincenzo~Andretta \and Thierry~Appourchaux \and Fr\'ed\'eric~Baudin \and Luis~Bellot-Rubio \and  Aaron~C.~Birch \and Patrick~Boumier \and Robert~H.~Cameron \and Matts~Carlsson \and Thierry~Corbard \and Jackie~Davies \and Andrew~Fazakerley \and Silvano~Fineschi \and Wolfgang~Finsterle \and Laurent~Gizon \and Richard~Harrison \and Donald~M.~Hassler \and John~Leibacher \and Paulett~Liewer \and Malcolm~Macdonald \and Milan~Maksimovic \and Neil~Murphy \and Giampiero~Naletto \and Giuseppina~Nigro \and Christopher~Owen \and
\mbox{Valent\'in~Mart\'inez-Pillet} \and Pierre~Rochus \and Marco~Romoli \and Takashi~Sekii \and Daniele~Spadaro \and Astrid~Veronig \and W. Schmutz 
}

\institute{L. Harra \at
             PMOD/WRC,  Dorfstrasse 33, CH-7260 Davos Dorf  and ETH-Z\"urich,  Z\"urich, Switzerland \\
              \email{louise.harra@pmodwrc.ch}; ORCID: 0000-0001-9457-6200 
              \and V. Andretta \at INAF, Osservatorio Astronomico di Capodimonte, Naples, Italy \email{vincenzo.andretta@inaf.it}; ORCID: 0000-0003-1962-9741     
              \and T. Appourchaux \at Institut d'Astrophysique Spatiale, CNRS, Université Paris--Saclay, France; \email{Thierry.Appourchaux@ias.u-psud.fr}; ORCID: 0000-0002-1790-1951
              \and F. Baudin \at Institut d'Astrophysique Spatiale, CNRS, Université Paris--Saclay, France; \email{frederic.baudin@universite-paris-saclay.fr}; ORCID: 0000-0001-6213-6382
              \and L. Bellot Rubio \at  Inst. de Astrofisica de Andalucía, Granada Spain
              \and A.C. Birch \at Max-Planck-Institut f\"ur Sonnensystemforschung, 37077 G\"ottingen, Germany; \email{birch@mps.mpg.de}; ORCID: 0000-0001-6612-3861
              \and P. Boumier \at Institut d'Astrophysique Spatiale, CNRS, Université Paris--Saclay, France;  \email{patrick.boumier@ias.u-psud.fr}; ORCID: 0000-0002-0168-987X
              \and R. H. Cameron \at Max-Planck-Institut f\"ur Sonnensystemforschung, 37077 G\"ottingen, Germany; \email{cameron@mps.mpg.de}; ORCID: 0000-0001-9474-8447
              \and M. Carlsson \at Institutt for teoretisk astrofysikk Svein Rosselands hus Sem Sælands vei 13 0371 OSLO, Norway. 
              \and T. Corbard \at Universit\'e C\^ote d'Azur, Observatoire de la C\^ote d'Azur, CNRS, Laboratoire Lagrange, Nice, France; \email{thierry.corbard@oca.eu}; ORCID: 0000-0002-9615-9619
              \and J. Davies \at STFC RAL Space, Rutherford Appleton Laboratory, Harwell Campus, Oxfordshire, OX11 0QX, UK; \email{jackie.davies@stfc.ac.uk}; ORCID: 0000-0001-9865-9281
              \and A. Fazakerley \at UCL-MSSL, Holmbury St Mary, Dorking, RH5 6NT, UK
              \and S. Fineschi \at INAF, Osservatorio Astrofisico di Torino: Pino Torinese, Piemonte, Italy; ORCID: 0000-0002-2789-816X
              \and W. Finsterle \at PMOD/WRC,  Dorfstrasse 33, CH-7260 Davos Dorf, Switzerland
              \and L. Gizon \at 
              Max-Planck-Institut f\"ur Sonnensystemforschung, 37077 G\"ottingen, Germany; Institut f\"ur Astrophysik, Georg-August-Universit\"at G\"ottingen, 37077 G\"ottingen, Germany; Center for Space Science, New York University Abu Dhabi, Abu Dhabi, UAE; \email{gizon@mps.mpg.de}; ORCID: 0000-0001-7696-8665
              \and R. Harrison \at STFC RAL Space, Rutherford Appleton Laboratory, Harwell Campus, Oxfordshire, OX11 0QX, UK; \email{richard.harrison@stfc.ac.uk}; ORCID: 0000-0002-0843-8045
              \and D. Hassler \at Instrumentation and Space Research Division
Southwest Research Institute, Boulder, CO, USA
              \and J. Leibacher \at Institut d'Astrophysique Spatiale, CNRS, Université Paris--Saclay, France; Max-Planck-Institut f\"ur Sonnensystemforschung, G\"ottingen, Germany;
              Lunar and Planetary Laboratory, University of Arizona, Tucson, Arizona, USA;
              National Solar Observatory, Boulder, Colorado, USA; \email{john.leibacher@universite-paris-saclay.fr};
              ORCID: 0000-0001-7605-3684
              \and P. Liewer \at Jet Propulsion Laboratory
M/S 169-506, 4800 Oak Grove Drive, Pasadena, CA 91109 USA
              \and M. Macdonald \at University of Strathclyde, UK\\
              \email{malcolm.macdonald.102@strath.ac.uk}; ORCID: 0000-0003-4499-4281
              \and M. Maksimovic \at LESIA, Observatoire de Paris, Université PSL, CNRS, Sorbonne Université, Université de Paris, 5 place Jules Janssen, 92195 Meudon, France; \email{Milan Maksimovic <milan.maksimovic@obspm.fr}; ORCID: 0000-0001-6172-5062
              \and N. Murphy \at Jet Propulsion Laboratory
M/S 169-506, 4800 Oak Grove Drive, Pasadena, CA 91109, USA
              \and G. Naletto \at University of Padova, Italy; ORCID: 0000-0003-2007-3138
              \and G. Nigro \at University of Calabria, Arcavacata di Rende, Italy\\
              \email{giusy.nigro@fis.unical.it}; ORCID: 0000-0001-8044-5701
              \and C. Owen \at UCL-MSSL, Holmbury St Mary, Dorking, RH5 6NT, UK 
              \and V. Mart\'inez-Pillet \at National Solar Observatory, Boulder, Colorado, USA
              \and P. Rochus \at CSL, Belgium
              \and M. Romoli \at Arcetri, Italy; ORCID: 0000-0001-9921-1198
              \and T. Sekii \at NAOJ, Japan
              \and D. Spadaro \at INAF, Osservatorio Astrofisico di Catania, Catania, Italy; \email{Daniele.Spadaro@oact.inaf.it}; ORCID: 0000-0003-3517-8688
              \and A. Veronig \at University of Graz, Austria
               \and W. Schmutz \at PMOD/WRC, Davos, Switzerland
}
 
 \date{Received: date / Accepted: date}

\maketitle
\hbadness=60000
\begin{abstract}
A mission to view the solar poles from high helio-latitudes (above 60$^\circ$) will build on the experience of Solar Orbiter as well as a long heritage of successful solar missions and instrumentation (e.g. SOHO \cite{SOHO}, STEREO \cite{stereo}, Hinode \cite{Hinode}, SDO \cite{SDO}), but will focus for the first time on the solar poles, enabling scientific investigations that cannot be done by any other mission. One of the major mysteries of the Sun is the solar cycle. The activity cycle of the Sun drives the structure and behaviour of the heliosphere and is, of course, the driver of space weather. In addition, solar activity and variability provides fluctuating input into the Earth climate models, and these same physical processes are applicable to stellar systems hosting exoplanets. One of the main obstructions to understanding the solar cycle, and hence all solar activity, is our current lack of understanding of the polar regions. In this White Paper, submitted to the European Space Agency in response to the Voyage 2050 call, we describe a  mission concept that aims to address this fundamental issue.

In parallel, we recognise that viewing the Sun from above the polar regions enables further scientific advantages, beyond those related to the solar cycle, such as unique and powerful studies of coronal mass ejection processes, from a global perspective, and studies of coronal structure and activity in polar regions. Not only will these provide important scientific advances for fundamental stellar physics research, they will feed into our understanding of impacts on the Earth and other planets' space environment.

\end{abstract}

\section{Introduction}
\label{intro}

It has long been a scientific goal to study the poles of the Sun, illustrated by the NASA/ESA International Solar Polar Mission that was proposed over four decades ago, which led to ESA's Ulysses spacecraft \cite{Ulysses} (1990 to 2009). Indeed, with regard to the Earth, we took the first tentative steps to explore the Earth's polar regions only in the 1800s. Today, with the aid of space missions, key measurements relating to the nature and evolution of Earth's polar regions are being made, providing vital input to climate-change models. The poles of other planets have also been explored using spacecraft, revealing the ice caps of Mars and the unexpected and intriguing polar flow patterns of the gas giants Jupiter and Saturn. However, the polar regions that remain largely uncharted are the poles of our Sun, yet it is those polar regions that hold the key for our understanding of the activity cycle of our star and, thus, of its impact on the environment of all planets in our Solar System, including our own.

Ulysses used a Jupiter fly-by to achieve an orbit with an inclination 80$^\circ$ out of the ecliptic plane, and, from distances as close as 1.34 AU was able to measure the in-situ particle and field environments above the Sun's polar regions; Ulysses did not carry instrumentation to image the Sun. However, the mission did provide seminal observations of different particle and field environments of the solar wind emerging from polar and equatorial regions. That said, to date we are blind as to what the solar poles actually look like and how they behave in enough detail to understand the role that the polar regions play in, for example, the solar cycle.

In 2020, ESA's Solar Orbiter mission \cite{orbiter2020} was launched, with an array of 10 instruments to probe directly the solar wind, as previous missions have done, but also directly image the sources of the solar wind. Solar Orbiter is a solar encounter mission, with prime orbits allowing solar approaches to within Mercury's orbit every 150 days. The later phases of the mission use Venus gravity assists (VGA) to slowly raise the spacecraft orbit out of the ecliptic plane to reach 30$^\circ$ helio-latitude. The mission's focus is the study of links between solar wind measurements from the in-situ instrumentation and sources of the solar wind from the remote sensing instrumentation. This is the first time that remote sensing of the sources of the solar wind has been done hand in hand with in-situ measurements this close to the Sun. The out-of-ecliptic phase will provide the first ever view of the poles, towards the end of the mission lifetime. 

This paper builds on the knowledge that will be gained with Solar Orbiter science, and it describes a mission with the principal focus being the exploration of the poles -- for long durations. We describe the scientific drivers behind this mission and the technical challenges that lie ahead.

\section{Science goals}
\label{sec:science}
There are four key science goals that this mission should address. These are:
\begin{itemize}
\item To study the interior of the solar polar regions to uncover the key role of magnetic flux transport in the solar cycle
\item To study the global mass-loss of a star through discrete mass ejection processes
\item To determine solar irradiance at all latitudes
\item To explore solar activity at the poles and the impact on the solar wind
\end{itemize}
We will describe these goals in the following sub-sections. For each science goal we will describe the requirements necessary to fulfil the science. 

\subsection{Science goal 1: To study the interior of the solar polar regions to uncover the key role of magnetic flux transport in the solar cycle. }
We are woefully ignorant of the large-scale circulation of plasma in the Sun and stars and how it interacts with stellar rotation and convection to generate i) the large-scale long-timescale (22 years) solar magnetic (sunspot number) cycle and ii) small-scale and short-timescale (months) appearance and evolution of magnetic active regions. For the Sun, where we can observe these flows at low helio-latitudes ($<$ 60$^\circ$), we have the beginnings of an observationally motivated model of how the magnetic field is generated by the dynamo process.

On the theory side, we have various increasingly sophisticated dynamo models, and we have an improving suite of numerical simulations that are becoming progressively more realistic. However, what we are missing is an understanding of how magnetic flux interacts with rotating turbulent convection to give rise to the solar cycle below the solar surface and to form flux concentrations that then lead to the observed active regions on the surface of the Sun. 
By observing the Sun at high latitudes where the poleward flows converge, we will learn about the effects of rotation on convection and their combined effects on the magnetic field.

The evolution of the magnetic flux is closely tied to the latitudinal differential rotation of the Sun and displays a complex dynamic behaviour. In addition to the systematic decrease in the rotation rate towards the pole, rotation varies through the solar magnetic activity cycle (the ``sunspot cycle") at most latitudes (the ``torsional oscillation''). The magnetic flux is dragged by the poleward meridional flow, which sets the duration of the cycle in some dynamo models.  The observed poleward flows simply cannot continue to transport mass to the poles; mass cannot accumulate there without limit:  there must be a (to date unobserved) return flow toward the equator within the Sun to conserve mass \cite{howelrsp,choudhuri2021}.

\begin{figure}
      
\includegraphics[scale=.56]{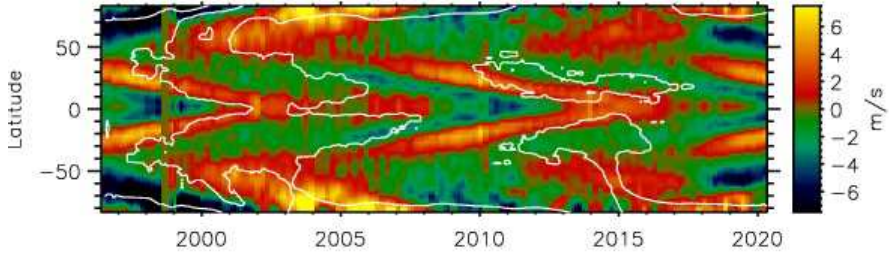}
\caption{Zonal flow  residuals, or torsional oscillations,  1\,\% of the solar radius below the visible surface from MDI and HMI f-mode splittings (adapted from \cite{corbard}). The white contours  show the 2 Gauss limit of the unsigned radial magnetic field averaged over longitudes for each Carrington rotation covered by MDI and HMI magnetograms (see also Fig.~\ref{fig:6}).}
\label{fig:1} 
\end{figure}

Large-scale internal solar flows and differential rotation can be measured at low and mid helio-latitudes by the helioseismology of globally resonant sound waves \cite{jcd2001} while the local helioseismology of propagating waves \cite{gizon} can measure the  variations of the larger scale, complex, temporally varying flows at low helio-latitudes. 
These studies have demonstrated the enormous power of these techniques at low latitudes where we currently have access. Helioseismology has revolutionized our views of the structure and dynamics of the convective region and the solar dynamo \cite{cameron,charbonneau}. In particular, helioseismology has revealed the differential rotation in the solar interior (e.g. \cite{thompson}), a key ingredient in dynamo models. Helioseismology also revealed the depth-dependence of the time-varying zonal flows, as well as the near-surface meridional circulation, constraints on the amplitudes of convective velocities, and inflows around active regions -- all of which have improved our understanding of the advection and stretching of the magnetic field through the solar cycle. These flows vary over the longest time scales and are different from cycle to cycle (Figure~\ref{fig:1}); they are important probes of the global dynamics, including the dynamics of the magnetic field. 

Torsional oscillations are variations of the solar  differential rotation that present two distinct branches: a low-latitude branch that progresses over time from about $50^\circ$ toward the equator as sunspots do and a polar branch that progresses from $50^\circ$ toward the poles as diffuse magnetic fields do (Figure~\ref{fig:6}). Both branches are thought to be strongly linked to the magnetic cycle (Figure~\ref{fig:1}) but possibly through different physical processes \cite{rempel2007}. The high-latitude branch has been found notably less strong in Cycle 24 than Cycle 23. Assuming that this poleward branch is driven through Lorentz force feedback, \cite{rempel2012} found that the transition of the dynamo toward a lower state of magnetic energy is linked to a drop in the high latitudes' rotation rate leading to the apparent fading of the polar branch. In order to investigate this link between zonal flows and the magnetic field near the pole, it is thus important to get a better resolved view of both quantities at high latitudes.
It is essential that these techniques be applied  near the poles, where circulation patterns must return to the equator within the the interior, where the solar cycle begins, and where the high-speed solar wind escapes into the heliosphere.
An emerging field of study is energy transport by global-scale waves (e.g. Rossby waves) and convective modes (e.g. giant cells); this is made possible by very long duration observations (many years) and should be extended over multiple cycles and particularly at high latitudes
\cite{loeptien2018,liang2019,Hathaway2020}.

The first indications of a new solar magnetic cycle occur at high latitudes where they are difficult to observe from the ecliptic. Figure~\ref{fig:1} shows the current observations of the magnetic field and the flows, which reach only to 70$^\circ$. These high latitude emergences are more direct probes of the newly wound-up magnetic field than the low latitude active regions that are built up over many years. The out-of-ecliptic views will also enable us to map flux emergences on the farside and study non-axisymmetric dynamo modes. The same views will enable the study of the polar  magnetic field reversals in detail.

\begin{figure}
  \includegraphics[width=\textwidth]{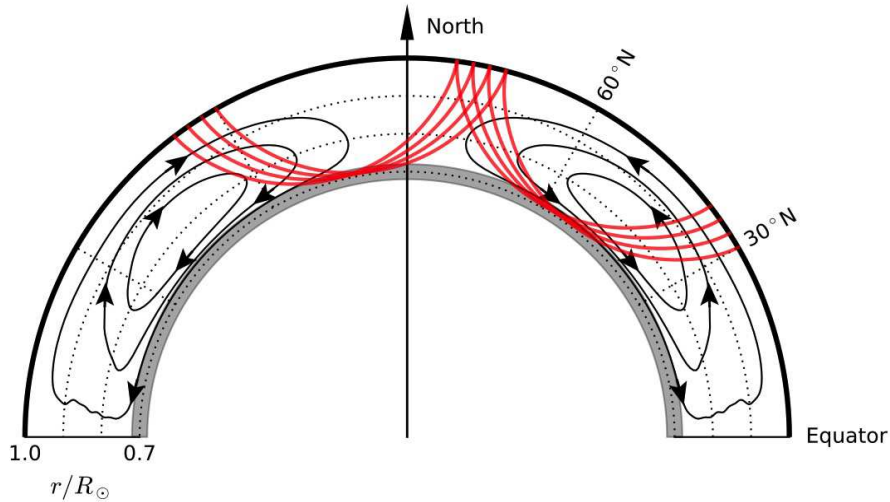}
\caption{Example p-mode ray paths (red curves) accessible to a solar polar mission. These ray paths could be used to measure the subsurface meridional flow at high latitudes. The meridional flow stream lines shown here in the northern hemisphere (black curves with arrows) were inferred using time--distance helioseismology and GONG observations  over Solar Cycle 24  \cite{Gizon2020}. The region in grey highlights  the base of the convection zone. [Image credit Zhi-Chao Liang.]}
\label{fig:2}       
\end{figure}

\subsubsection{Science goal 1: requirements}

Technically, the helioseismic science objectives that drive a solar polar mission's requirements are orbit, data rate, and the total duration of the observations. Helioseismology requires nearly continuous, full-disc observations at a temporal cadence of roughly one minute in order to resolve the entire acoustic spectrum. The fundamental scales of solar dynamics and activity define two basic observing modes that are needed to achieve the helioseismology objectives: 
(1) relatively short (7\,--\,14 days) and high-resolution (500 $\times$ 500 km) Dopplergrams of sound waves and intensity images of individual convective granules, and 
(2) multiple intervals of long time intervals (36\,--\,72 days); at lower spatial resolution (6000~km $\times$ 6000~km) to investigate global dynamics with helioseismic techniques.
The longest runs are needed to achieve the frequency precision to resolve the tachocline structure and dynamics. High-latitude viewing of solar p-modes is not only necessary to study the dynamics in the polar regions, it is also essential for studying the deep convection zone, including longitudinal structures in the tachocline (see Figure~\ref{fig:2}). An orbit with 60$^\circ$ inclination would allow determination of the flows in the polar regions of the upper convection zone. In addition, the high-latitude orbit will allow us to perform stereoscopic helioseismology, to extend the regions throughout the solar interior that are accessible.
To achieve these objectives, the total mission duration should be several years.

\subsection{Science goal 2: To study the global mass-loss of a star through discrete mass ejection processes}
This science goal recognises the unique advantages of studying the processes giving rise to the mass loss from the Sun, in particular coronal mass ejections (CMEs), from a high-latitude perspective. Such an observational vantage point would enable an unprecedented opportunity for the study of the global mass loss from a star.

\begin{figure}
  \includegraphics[scale=1.35]{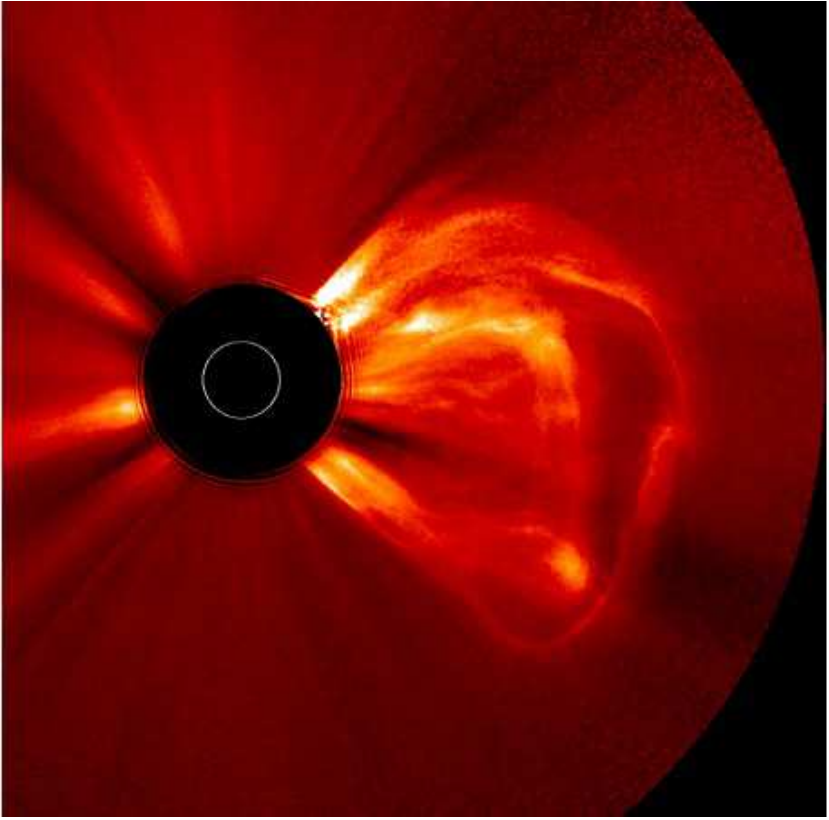}
\caption{A CME detected using the STEREO/COR2 coronagraph in July 2011. Historically, CME observations have been made from the ecliptic plane. Geometrical considerations mean that CME observations from high helio-latitude would provide unique opportunities to study solar global mass loss with a 360$^\circ$ view with respect to the ecliptic plane. [Image credit NASA/COR2 team, see STEREO Gallery at \url{https://stereo.gsfc.nasa.gov/gallery/gallery.shtml}]}
\label{fig:3}       
\end{figure}

CMEs, large-scale eruptions of hot plasma that may accelerate charged particles and can travel well beyond Earth's orbit, were discovered in the early 1970s using the space-borne coronagraphs aboard OSO-7 \cite{tousey1973} and Skylab \cite{koomen1974}.
Since this time, a number of space missions have provided a wealth of coronagraph observations of CMEs (e.g. Figure~\ref{fig:3}), all from the ecliptic plane, and all but one, the Solar Terrestrial Relations Observatory (STEREO), from a near-Earth vantage point; the twin STEREO spacecraft orbit the Sun and, although observing from near the ecliptic plane, make observations from off the Sun-Earth line.

Many studies have shown that CMEs originate from the solar activity belts, defined by the location of active regions. The activity belts generally occupy restricted regions below 30$^\circ$ latitude but migrate equatorward with decreasing activity. Statistical analysis of 7000 CMEs imaged by the coronagraphs on the near-Earth SOHO spacecraft \cite{yashiro} clearly demonstrated the latitudinal variation of CMEs with the solar cycle. In addition, CMEs often show an equatorward deflection after eruption, resulting in a greater concentration near the equatorial streamer belt. 

Coronagraphs detect CMEs through Thomson scattering of photospheric white-light off free electrons in the CME structure. Coronagraphs are more sensitive to CMEs that are not directed towards the spacecraft, although the densest spacecraft-bound CMEs can be detected as faint ``halo" events emerging from behind the coronagraph occulting discs. Thus, from a polar platform, taking advantage of the observational geometry and the confinement of CMEs to the activity belts, for the first time, we have a vantage point from which we can potentially detect all CMEs effectively, and especially those Earth directed, and provide a global overview of CME activity of the Sun. 

This would not only provide oversight of mass ejection phenomena in terms of global distribution and frequency, it would also permit the investigation of the longitudinal structure and density distribution of the activity belts in the corona, as well as providing new insights into the kinematic and topological parameters of CMEs, including their complex magnetic structure. Such unique and powerful studies of CME processes from a polar perspective are therefore crucial for a deeper understanding of impacts on the Earth and other planets' space environments. Recently, \cite{xiong2018} synthesized the white-light emission of an Earth-directed CME using a 3D MHD code and derived its structure as observed from out of the ecliptic plane, demonstrating the feasibility of the investigations quoted above.

\begin{figure}
  \includegraphics[scale=.8]{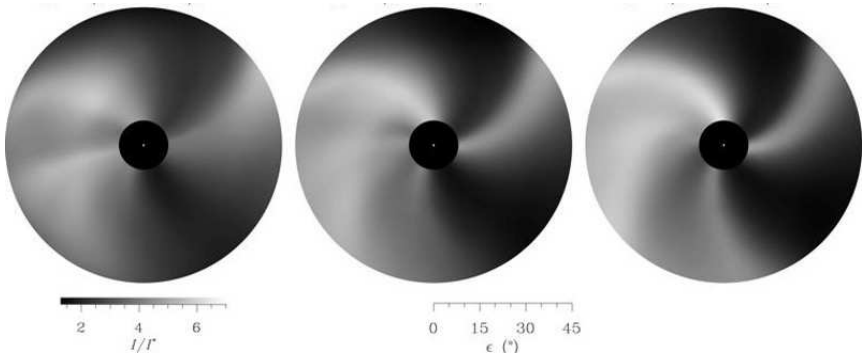}
\caption{Simulated images of the CIR structure observed from out of the ecliptic plane \cite{xiong2017}. The images show normalised white-light brightness (I/I*) for elongations ($\epsilon$) out to 45 degrees. The concentric black and white circles correspond to the inner elongation and the solar-disc size respectively.}
\label{fig:4}       
\end{figure}

In addition to CME studies, the polar vantage point provides an opportunity to observe, for the first time, the background corona/inner heliosphere from high-latitudes, enabling complete mapping of the longitudinal distribution over 360 deg and temporal evolution of the equatorial streamer belt over prolonged periods.  These observations would also facilitate the derivation of the mass and energy flux carried away by the solar wind in the solar equatorial plane. Observations from over the pole would not only permit us to investigate outward-propagating CMEs, we will also be able to image the development of co-rotating interaction (CIR) regions in the background solar wind. These interaction regions are curved because the magnetic field lines that define their topology are curved due to solar rotation.
The occurrence and structure of CIRs have been identified from images from the Heliospheric Imagers aboard STEREO \cite{alexis}. Whilst this is an important step, such images are still restricted to equatorial regions; a polar vantage point would enable a far superior view of CIR global structure and evolution, and interaction with the Earth and other planets. \cite{xiong2017} synthesized the out of ecliptic view (Figure~\ref{fig:4}), demonstrating the possibilities of imaging the 3D CIR structure from above.           

The occurrence rate of CMEs strongly varies over the solar cycle \cite{webb2017}. During solar maximum, the average CME rate is about 5 per day, and multiple CMEs can occur in close succession associated with the same active regions (homologous and sympathetic eruptions). This inevitably leads to the interaction of successive CMEs either close to the Sun or in interplanetary space. Such interaction is associated with many complex processes -- momentum exchange, magnetic reconnection, and propagation of magnetosonic shock waves through the ejecta, e.g. \cite{lugaz,manchester}.
These phenomena also cause changes of the CME structure, such as its radial extent, expansion speed and field strength, which in turn affect its capacity for causing a geomagnetic disturbance. The most intense geomagnetic storms are associated with interacting CMEs propagating in the ecliptic \cite{vennerstrom,liu}. CME geo-effectiveness can also be enhanced when the CME is compressed from behind by a CIR or high-speed solar wind stream, resulting in enhanced magnetic fields (e.g. \cite{kilpua}). A polar perspective will provide a much better view of CME--CME and CME--CIR interaction processes in interplanetary space, where they occur and how they propagate.

Viewpoints above the poles can allow a more complete determination of the fractional contribution of CMEs to the mass loss of the Sun, particularly in synergy with complementary in-situ measurements. These are very useful for the study of solar mass loss, including steady solar wind and transients events.  The cumulative effect of this transient mass loss in the form of CMEs could be larger than current estimates, typically $<$ 10\% of the overall solar wind mass flux (about 2$\times$10$^{-14}$ M$_{\odot}$/yr). 

Mass loss via the solar wind and CMEs can be investigated as a function of solar magnetic activity, with the purpose of predicting the CME occurrence rate and associated mass loss rate of solar-type stars on the basis of their magnetic activity level \cite{mishra}. 
The relative contribution of stellar CMEs into the stellar wind could be high for active stars \cite{aarnio2011}; quantification of this is important in evaluating the global mass-loss from  the stars.
Stellar mass loss is also found to have a significant influence on stellar evolution since this may determine stellar spin down, with the consequential impact that rotation plays on stellar properties. In fact, it is widely accepted that the Sun, like other late-type stars that have overcome the disc-stellar interaction phase, undergoes loss of angular momentum during its main sequence lifetime mostly due to wind magnetic braking \cite{kraft,Mestel}. This is a mechanism generated by the stellar wind torque \cite{lanzafame,matt,finley}, due to the fact that ejected plasma remains magnetically connected to the stellar surface for several stellar radii, namely up to the Alfv\'enic radius estimated to be around 15\,--\,25 R$_\odot$ for the Sun \cite{kwon}. The same mechanism of wind magnetic braking has been proposed for CMEs \cite{aarnio2011}; the mass loss due to the cumulative effect of all CMEs should have a contribution to the total torque that spins down the  star. 
This highlights the importance of determining the total mass-loss of the Sun, considering also the cumulative effect of all CMEs (including those observationally not yet resolved), also from the perspective of making predictions about the future evolution of our star.

Difficulty in evaluating 
solar and stellar wind torques arises mostly due to the lack of observations (past, current, or planned) of mass-loss and stellar wind on a global scale and of the magnetic field geometry of these stars and their topological magnetic interconnection with their wind. The only hope for gaining some indication is by observing our closest star, and extending this to other late-type stars 
in the framework 
of the solar-stellar connection \cite{brun2015,brun2017}. To assess angular momentum loss rate, the magnetic topology of the star and, more specifically, the complexity of the surface magnetic field at all latitudes has been shown to be crucial in the coupling between a star and its wind \cite{garraffo}. Moreover, the global coverage of magnetic field measurements including also the poles, is expected to substantially improve current MHD models of the solar wind and CME propagation through interplanetary space. 
A viewpoint from above the solar poles, will provide huge benefits, for example avoiding (1) the projection effects on the line of sight magnetic field measurements that produce a large amount of noise at the poles when polar regions are observed from low latitudes near the ecliptic, and (2) the Sun's tilt angle periodically rendering those areas invisible to current and planned solar probes (i.e. Parker Solar Probe and Solar Orbiter). The need for a mission like the one we are proposing is also critical, considering that these kinds of measurements are not possible, not even in the near future, for late-type stars other than the Sun.

Overall, it would be extremely valuable to compare CME events as seen from a polar view with similar instruments deployed in the ecliptic plane from other space science missions that are operational (e.g. space weather monitoring spacecraft at L1 and L5, carrying coronagraphs, EUV imagers, and heliospheric imagers). The successful scientific outcome of a dedicated solar polar mission would not be dependent on the coincidence of solar missions in the ecliptic plane, but, using such observations in concert would lead to significant additional advances. 
Such combined observations would enable analyses of CME phenomena in both longitude and latitude for the first time; 3D views of CMEs (and indeed CIRs) would be extremely important for studies of CME onset, propagation, and impact studies. Exploiting available spacecraft combinations would also give the opportunity for investigating the 3D global structure of the solar corona and heliosphere, and how this is influenced by solar activity and, in particular, CMEs. This new, combined observational approach would contribute significantly to the validation of global heliospheric  models. 

\subsubsection{Requirements for Science Goal 2}

The concept of a polar mission providing extensive coverage of the inner helio\-sphere, focused on the ecliptic plane, for extended periods of time, would be unique, and the scientific advantages of such a mission have been spelt out above. Our goal is to achieve: 

\begin{itemize}
\item	helio-latitudes of greater than or equal to 60$^\circ$; 
\item	for many periods of tens of days each at the poles, across a long portion of a solar cycle;
\item	mission orbital parameters in terms of aphelion and perihelion values well within 1 AU, dictated by the required resolution of the remote sensing instrumentation.
\end{itemize}

The high-latitude view would allow crucial measurements of the directionality and changes of speed of the plasma erupted from the Sun, and unique tracking capabilities of events with respect to all Solar System bodies. If the spacecraft scientific payload included a coronagraph and a heliospheric imager, working in concert, the observation and tracking of CMEs and CIRs and their influence throughout the corona and inner heliosphere, over all longitudes, would be available for the first time. This places some requirements on the angular extent of the combined field of view (coronagraph plus heliospheric imager, without any gap): from 1\,--\,2 deg (inner edge of the coronagraph) to about 45 deg (outer edge of the heliospheric imager) from the centre of the Sun.
The 45 deg limit assumes a distance of the spacecraft from the Sun of 1\,AU, enabling the Earth to appear at the edge of the field of view. This can be considered to be a minimum requirement. With an additional EUV imager and magnetograph the mission would include the magnetic and coronal imaging capabilities required to investigate the solar sources of the solar wind phenomena. Finally, the inclusion of particle and field in-situ instrumentation would allow measurements of the local plasmas to provide the in-situ `ground truth' measurements relevant to several of the lines of research mentioned above.

\subsection{Science goal 3: To determine the solar irradiance at all latitudes}

The total solar irradiance (TSI) has been measured since 1978 by a number of space instruments. All TSI space missions so far have been bound to the ecliptic, which coincides closely with the solar equatorial plane. While the ecliptic plane is the perfect vantage point for monitoring the essential climate variable TSI, in other words the energy input on planet Earth, a polar mission will be able to explore how the Sun's radiative output varies at all latitudes which ultimately allows the determination of the solar luminosity. We know that the TSI varies as a function of the solar activity cycle. The key driver of the solar activity cycle is understood to be the solar surface magnetic field which manifests itself as the dark sunspots, and bright faculae and network. The latter compensate for the dark sunspots. The polar regions remain a mystery, and the dominant source of radiative output at the poles is not well understood. Indeed, the interaction between the poles and the activity belt and how that changes with the cycle is not clear. Another interesting question is the long-term changes in the solar cycle -- the past two cycles of the Sun have been weaker than the previous ones. Measuring the solar irradiance during weak cycles at the minimum period may provide insight into the potential long-term minima such as the Maunder minimum -- what actually causes a reduction in the radiative output?
This polar mission concept will measure the TSI and alongside that telescopes will observe the bright features on the Sun. Measurements from above the solar equator only provide a first-order estimate of the total energy output of the Sun, the solar luminosity. 

A TSI radiometer on board a solar polar orbiting mission would be able to measure the latitudinal distribution of the solar irradiance and how it varies with time. Accurate measurements of the solar luminosity will feed in to the understanding of solar-type stars whose orientation of the rotation axis is unknown, i.e. it is not known from which vantage point the stars are observed. In recent years, however, asteroseismology for example, has revealed that differential rotation also occurs on Sun-like stars (e.g. \cite{Benomar}). This study did show that this differential rotation can be larger in other stars than it is in our Sun. Understanding how the differential rotation on our Sun impacts the irradiance will be key in understanding the luminosities of other stars with different differential rotation, and different orientations.

\begin{figure}
  \includegraphics{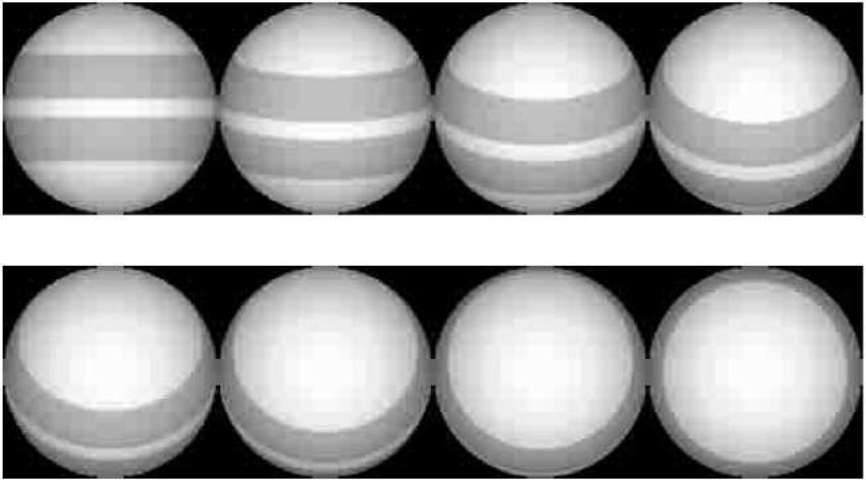}
\caption{Visibility of two zonal latitude bands between $\pm$5$^\circ$and $\pm$30$^\circ$ latitude for different inclinations $i$ (defined here as the angle between the solar rotation axis and the line-of-sight); $i$ decreases in steps of $\Delta i \approx$13$^\circ$ from the upper left ($i$ = 90$^\circ$) to the lower right ($i$ = 0$^\circ$): Adapted from \cite{Knaack}.}
\label{fig:5}       
\end{figure}

A key question is whether the brightness of the poles is different than the brightness of the quiet areas elsewhere on the Sun. \cite{Knaack} studied the influence of the solar inclination, $i$, on the outgoing total and spectral solar flux. With their model they found that the total flux increases by about 0.15\,\% when measured from the poles ($i$ = 0$^\circ$, see Figure~\ref{fig:5} bottom right) with respect to $i$=90$^\circ$ (upper left). The authors also find that, while UV variability decreases slightly when observed from the poles, variability in the visible is expected to increase by up to 150\,\%.   
If this model is confirmed, this means that the shape of the spectrum depends on the inclination angle of the solar (or stellar) observation. Ultimately, this would then also have implications on the stellar magnetic activity index (S-index) of both the Sun and Sun-like stars. Currently, the S-index of the quiet Sun, S$_\textrm{QS}$, is understood to be independent of the inclination. However, if the shape of the spectrum is indeed a function of $i$, so would be the S-index. For comparing the Sun with solar-type stars the S-index plays a crucial role and, therefore knowing its exact value is essential.
In summary, a dependence of the directional energy output of the Sun cannot be ruled out and its exact amplitude needs to be determined. This result would be crucial to constrain irradiance models for solar and stellar applications.

\begin{figure}
  \includegraphics[scale=.3]{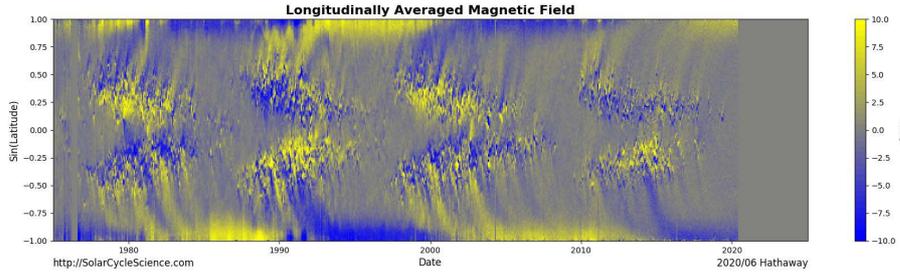}
\caption{ Solar magnetic cycle from 1974 to the present. Opposite polarity is indicated by the blue and yellow. [Image credit D. Hathaway]}
\label{fig:6}       
\end{figure}

The luminosity of the Sun is $L=4\,\pi$ (1\,AU)$^{2} S$, with the assumption that the TSI is a suitable measure of the flux $S$. The flux measurements from a polar mission will directly give us the information as to how much S varies as a function of latitude and ultimately whether $L$ varies over the solar cycle. Figure~\ref{fig:6} illustrates the solar magnetic cycle which covers on average 22 years. During this time the poles undergo a magnetic field reversal. We know that TSI varies with the solar cycle and these measurements will address whether the solar luminosity also varies over the solar cycle. We will be able to measure if the brightness of the poles change over the solar cycle and what role the polar magnetic field strength plays. 

With the search for habitable planets there is increasing need to understand the place of the Sun amongst other solar-type stars. However, differences exist between the Sun and Sun-like stars that are not yet fully understood. A striking difference is the smaller variability of the Sun on solar cycle time scales, e.g. \cite{Lockwood,Radick}, compared to Sun-like stars. This difference might be partly due to the inherently different solar and stellar observations. Until now, all TSI and solar spectral irradiance (SSI) observations have been obtained from the vantage point of Earth. Stellar observations, however, are typically obtained as chromospheric or photometric variability and from random stellar inclinations. The fact that stars are observed from different viewpoints might explain part of the difference in solar cycle variability. Indeed, \cite{Knaack} showed that in the visible wavelengths, bright faculae are less effectively compensated by the sunspots if the Sun was observed at higher latitudes. This indicates that the comparison between the solar and stellar variability is rather complex. Therefore, to better link solar versus stellar variability it is essential to measure the solar flux from all viewpoints, most importantly from high latitudes. Ultimately, by better understanding solar and stellar variability we will be able to better constrain future solar variability and at the same time improve the characterization of solar-like host stars of Earth-like exoplanets. 

Our understanding of the changes of the radiative energy output of the Sun is based on the modeling of the radiation emerging from the surface components of the Sun. These include -- besides the quiet, mostly non-magnetic Sun -- the solar activity features such as sunspots, faculae, and the network. The changing area contribution of these features leads to an overall variation of the outgoing radiative flux.
Our knowledge of the emergent radiation of the solar surface components stems from radiative transfer modeling of these components. A key basis for the radiative transfer modeling is semi-empirical atmosphere structures that are developed to reproduce spectral observations at intermediate spatial resolution \cite{fontenla1999,fontenla2009}. Depending on the spectral wavelength of interest, they are then used with different radiative transfer codes \cite{Ermolli,margit} to model the emergent spectrum. Recently, the radiative transfer modeling has been further extended to employing 3D MHD models \cite{Yeo}.
There is however an important missing link. Due to the lack of observations, the detailed characterization of the solar atmosphere at the poles has so far not been possible. In particular, it is not known whether the photosphere at the poles is darker (cooler) or brighter (hotter) than the average quiet Sun. Moreover, we do not know the detailed properties of the polar plasma, i.e. its temperature structure, density, flow patterns. In order to infer these properties, it would be essential to observe the poles with intermediate to high spectral resolution. Based on the observed spectra it would then be possible to derive the atmospheric structure of the poles. Detailed observations of the poles will allow us then to improve irradiance reconstruction models -- further constrained by global flux measurements -- and ultimately to better link solar with stellar observations.

\subsubsection{Requirements for Science Goal 3}

The science goals for the latitudinal variation of total solar irradiance require as high an inclination as is possible (in an ideal world 90$^\circ$, but anything $>$60$^\circ$ would be a significant advance on existing/planned observations). Due to the importance of long-term measurements, it is necessary to collect observations over several years -- ideally over a solar cycle. 

\subsection{Science goal 4: To explore solar activity at the poles and its impact on the solar wind}

Ulysses (1990\,--\,2009) undertook the only previous exploration of the heliosphere outside the ecliptic plane, making in-situ solar wind plasma, electromagnetic fields, and composition measurements. These measurements demonstrated the prevalence, particularly near-solar minimum, of fast, relatively uniform, solar wind at high northern and southern latitudes \cite{McComas1998,mccomas2003}, while more variable ``slow'' solar wind originates from the coronal streamer belt at low-latitudes. During solar minimum, the fast solar wind most likely originates from large polar coronal holes and subsequently expands with distance from the Sun to fill much of the hemispheric cavity, while the slow solar wind is more confined around the equatorial plane.  Ulysses also demonstrated that the fast solar wind contains `open' magnetic flux, where the spacecraft is connected directly along magnetic field lines and thus provides the most direct physical connection to the solar wind origin in the solar atmosphere (e.g. \cite{Cranmer}).  Although Ulysses polar observations were made at very high solar latitudes ($\approx$\,80$^\circ$), they were confined to a relatively large distance range (between 1.3 and 5.4 AU) from the Sun, due to the reliance on a Jovian gravity assist maneuver (GAM) to reach high latitudes.

The out-of-the-ecliptic solar wind will also be sampled in situ by the Solar Orbiter mission. Like Ulysses, Solar Orbiter carries instruments to measure the solar wind plasma, electromagnetic fields, energetic particles, and composition, although significantly more modern variants. These measurements will be augmented by remote-sensing instruments providing imagery and spectroscopic information of the solar atmosphere. Solar Orbiter will make its measurements between 0.28 and $\approx$\,1\,AU.  Thus, although the mission will not reach the high latitudes sampled by Ulysses, it will sample the solar wind much closer to its sources.

There are strong scientific reasons to make solar wind measurements from a platform that combines both the high-latitude vantage point of Ulysses with the near-Sun, multi-instrument capabilities of Solar Orbiter. In the 1970s the Helios mission \cite{Rosenbauer} demonstrated that the sampling of pristine solar wind, free from the effects of in transit processing, requires in-situ measurements within a few tenths of an AU, a view confirmed through more recent imaging of the corona and inner heliosphere by, e.g., instruments on STEREO \cite{stereo}, and also also by the early results from the Parker Solar Probe mission \cite{Fox}.  A further advantage of sampling solar wind from high-latitudes would be extended periods of measurement of the expanding fast solar wind in the absence of any stream interactions with the slower wind from the streamer belt. Open magnetic flux in this region provides a persistent and direct magnetic connection between the surface magnetic field and ejected plasma elements. This can be modeled using magnetograph data, which can shed light on the acceleration of the solar wind and energetic-particle events.  Such measurements are crucial to providing constraints on solar wind formation processes on both a local and global scale and on the influence of the magnetic field close to the Sun. In particular, these measurements could be used to better determine how the spatial distribution of the fast solar wind properties varies with respect to the location/size/edge of the polar coronal holes. They would also help to determine the overall magnetic flux budget within the heliosphere, potentially resolving the ``open flux'' conundrum within existing measurements \cite{Linker}. These topics are expanded on further below.

{\bf (i) Identification of fast solar wind sources and acceleration.}

The Solar Orbiter mission philosophy recognises that direct, simultaneous measurements of both on-disc activity and in-situ features observed in the solar wind are needed to properly explore solar wind sources and related structures, such as polar plumes associated with polar coronal holes and the ambient solar wind from coronal holes.  Launching such a combined package on a solar polar orbiting platform reaching very high latitudes would complete these studies at all latitudes. As noted, measurements from such high polar latitudes would enable study of the evolution of the solar wind from its source region to the observing platform with little or no effects of, e.g., the stream interactions that significantly influence dynamics of the solar wind at lower latitudes. Such pristine fast solar wind observations will enable us to unambiguously establish their connection to features observed in the polar corona, such as polar plumes and coronal hole jets.  In turn, this allows us to investigate driving mechanisms.  Moreover, such measurements of the high-latitude solar wind with modern instrumentation, would also fill in parameter space and augment studies of processes occurring within the solar wind itself that have been made nearer the ecliptic (Solar Orbiter) or at greater distances (Ulysses).  For example, the level of plasma wave activity, turbulence, internal heating, and other kinematics within pristine fast solar wind from the polar coronal holes would provide a significant context in which to place similar measurements of more disturbed near-ecliptic regions. This is key knowledge needed to complete our understanding of how energy flows from a star to its surrounding environment. 

Observations from a polar vantage point have significant scientific advantages when combined with observations made from or near the ecliptic. Multiple vantage points significantly enhance our ability to undertake studies of the coronal structures that are the source of the solar wind.

{\bf (ii) Global knowledge of the solar wind, its sources, and in-transit dynamics.}

In addition to analyses of specific high-latitude solar wind sources and the processes involved in the release of the solar wind, synoptic polar observations from near-Sun high-latitude vantage points will complete an exploration of measurement space that will underpin a step-change in understanding the large-scale structure of the heliosphere and the physical processes therein. Data from a solar polar orbiting platform will reveal the full radial and longitudinal evolution of the solar wind and transient structures propagating through it, providing critical information on the fundamental nature of our star and its environment. For example, this would reveal the magnetic connectivity map across large volumes of space, which can be used to constrain theories of the role of the magnetic field in solar and heliospheric dynamics. Full measurements of the amount of open magnetic flux in the heliosphere can only be made by including measurements outside of the ecliptic. Combined with direct measurements of polar magnetic fields, their boundaries, and associated solar-wind source regions at high latitudes at the Sun, these out-of-ecliptic measurements will resolve inconsistencies in our understanding of the open flux. Such data can also be used to probe the global nature of key boundaries between the Sun and the heliosphere, such as the Alfv\'enic surface, and determine their roles in conditioning the outflowing solar wind plasma. Combined with ecliptic vantage points, the global nature of the interactions between solar wind streams with different sources, with different speeds, densities, etc., can be comprehensively studied.  In particular, the way these interactions evolve with both time and distance can be disentangled, which is very challenging based on measurements from one vantage point only.  The formation and evolution of CIRs and associated shocks, as well as CMEs, could in principle be tracked continuously over long periods of time from a polar vantage point (as discussed under Science Goal 2).  These measurements will provide a unique picture of the global variations in the fine structure within the fast solar wind, the latitudinal dependence of the occurrence of turbulence and wave--particle interactions, the influence of the three-dimensional structure of the magnetic field on these variations, the plasma convection and circulation flows at and below the surface, and how these might vary as a function of solar activity.  These are all important inputs for operational space weather forecasters.

Finally, we consider the propagation of energetic particles within the heliosphere. There is now a general consensus that solar energetic particles are generated by disparate processes, including direct release from solar flares and more gradual generation at interplanetary shocks propagating through the heliosphere in association with CMEs and/or CIRs. However, our understanding of the transport of these particles through the heliosphere remains rudimentary. STEREO observations show that energetic particle events apparently spread rapidly beyond regions that are magnetically connected to their source \cite{SEP}. This surprising result suggests that there must be processes operating near the Sun that act to not only accelerate the particles to their high-energies, but also transport them in radius, longitude and, presumably, latitude. Observations from a polar vantage point would be able to confirm the latter, but the relatively simple magnetic structure expected at higher latitudes will provide testable constraints on theories of these transport processes.

In addition, the propagation of energetic electron bursts through the inner heliosphere can be remotely sensed through observation of Type II and Type III radial emissions.  Their observation from various radial and latitudinal locations, or indeed from multiple vantage points in and out of the ecliptic, will enable better understanding of propagation of energetic electrons through space. Conversely, direct observation of these electrons can provide information on the magnetic connectivity of the spacecraft to solar sources and reveal the timings for the actions of the release processes in the corona.  This allows a more direct link with remote observations of potential source regions to be confirmed.
In addition to the specific advantages to making measurements of the Sun and the heliosphere from a solar polar vantage point, we note a more general benefit from such data in both priming and validating magnetic and plasma simulations. Magnetograph measurements of the solar poles would significantly augment the accuracy of the lower boundary conditions generally used in global simulations of the solar atmosphere and modeling of the heliosphere. Moreover, measurements of the solar wind distributed in radial distance, latitude, and longitude provide a network of ground truth points to which to compare and refine the simulation results (e.g. predicted vs observed flow speeds, densities, timings of magnetic reversals, etc.). These interdisciplinary activities to improve the accuracy of the model output would significantly benefit operational space weather services.

\subsubsection{Requirements for Science Goal 4}
The primary science of this goal will be achieved by an inclination of $>$\,60$^\circ$. This goal is the one mostly driven by the distance to the Sun -- it requires the location to be $<$1AU. A combination of remote sensing and in-situ instruments are required. 

\section{Space mission concepts}

There has been a drive for many years to explore the poles of the Sun. Table 1 summarises the range of ideas to date, and whether they fulfil the science goals stated above or not. The list may not be complete but is representative of the efforts to explore the solar poles and shows quite clearly that most (6 out of 8 of them) have not progressed beyond the concept stage -- mainly for technology readiness reasons. Only one has flown, namely Ulysses, which did not include the remote sensing payload required to address the goals described above. The remaining mission, Solar Orbiter, will achieve 32$^\circ$ helio-latitude, allowing some polar-oriented scientific studies towards the end of the mission but it was not designed to simultaneously and/or fully address all the goals presented here. To deliver the science that we present requires a new concept combining high latitude observations (above 60$^\circ$) with an appropriate orbit and remote sensing instrumentation.

\begin{table}
\resizebox{\textwidth}{!}{
\begin{tabular}{ l l l l l l l }
%

Mission & Inclination & Flight Status & SG 1 & SG 2 & SG 3 & SG 4 \\
\hline
Ulysses \cite{Wenzel}	& 80$^{\circ}$	& Flown	& no	& no	& no	& In-situ  \\
 & & & & & &only  \\
Solar Orbiter \cite{orbiter2020}  &	32$^{\circ}$ & Launched &	Short   &	Short   &	no & Short   \\
                                                     &   &     & duration & duration & & duration \\
SPDEx \cite{vourlidas2018} &	75$^{\circ}	$ & Concept &	yes &	yes &	yes &	yes \\
SPORT \cite{sport} & 	$\geq$ 60$^{\circ}$ & 	Concept &	Only fields, & yes	& no & 	yes \\
 & & & no helioseismology & & & \\
Polaris \cite{polaris}	& 75$^{\circ}$	& Concept &	yes &	yes &	yes &	yes \\
SPI \cite{liewer2008} & 	75$^{\circ}	$ &Concept &	yes	& yes &	yes &	yes \\
RAMSES \cite{ramses1998} & 	Above ecliptic &	Concept &	 no	& no &	no &	yes \\
European solar   &	90$^{\circ}$ &	Concept &	yes	& yes	& yes &	yes \\ 
STEREO mission \cite{bothmer1998} \\

\end{tabular}
}
\caption{A summary of missions that have been flown or planned to explore the poles. Only one polar mission has flown and it had no ability to remotely sense the poles. The ability of these missions to address each of the science goals (SG) presented in Section 2 is indicated.  }
\label{tab:1}
\end{table}


Each science goal described in this paper gives a summary of the requirements for the mission. These are based around the inclination from the ecliptic plane, and the duration of observing at the poles. These two main requirements will drive the technology of the mission concept. Table 2 summarises these requirements.

\begin{table}
\resizebox{\textwidth}{!}{
\begin{tabular}{l l l l l l }
Science  &	Inclination	& Distance 	& Duration	& Pole to	& Other \\ 
goal   &    &   to Sun &   & pole view? & \\ \hline
SG1& $>$60\,$^{\circ}$ &	$<$1AU	& $>$36 days at the poles & 	yes	& Ecliptic measurements beneficial \\
SG2	& $>$60\,$^{\circ}$	& $<$1 AU& Extended periods over solar cycle &	yes & 	Ecliptic imaging and in-situ beneficial \\
SG3	& $>$60\,$^{\circ}$ &	1AU &	Years (preferably solar cycle) &	yes &	Ecliptic TSI measurements beneficial \\
SG4	& $>$60\,$^{\circ}$ &  	$<$1 AU & 	10s days at the pole in each orbit &	no &	Requires remote sensing and in-situ
\end{tabular}
}
\caption{Key mission requirements for a polar mission}
\label{tab:2}
\end{table}

\subsection{Technology Challenges}

This section summarises the technical challenges of a mission to observe the poles, and how they might be achieved. A solar polar mission to exploit the scientific gains of observing the Sun and its near-space environment from high helio-latitudes must observe from above the poles for long durations. The goals, above, require a three-axis stabilised, solar-pointed platform carrying a package of remote sensing and in-situ instruments. The likely instruments would be an evolution of current instrument strengths on missions such as Solar Orbiter and STEREO. Here, we focus on the mission concept options with a view to the technology requirements. Several options would allow us to reach high helio-latitudes for long durations with increasing levels of technical difficulty.

\subsubsection{Ulysses type orbit with remote sensing instruments}

There has been one solar polar mission, the Ulysses mission. One option for a polar mission is to adopt an updated Ulysses-style orbital strategy, using a Jupiter gravity assist (JGA). Ulysses did not include solar remote sensing instrumentation. Observations were made over the polar regions only every 6 years in an orbit with aphelion 5.4 AU and perihelion 1.34 AU. A similar orbit, including remote sensing instruments, could achieve to some extent goals 2 and 3, but it is too restricted time-wise to achieve all of the scientific goals, and it is farther from the Sun than would be ideal for good spatial resolution imaging. Unlike Ulysses, we would require a three-axis stabilised spacecraft for the remote sensing instruments. In addition, a radioisotope thermoelectric generator (RTG) was used with Ulysses, given its distance from the Sun, and this may not be acceptable in the current climate.  

 \subsubsection{Solar sail option}
 
\begin{figure}
\includegraphics[scale=.28]{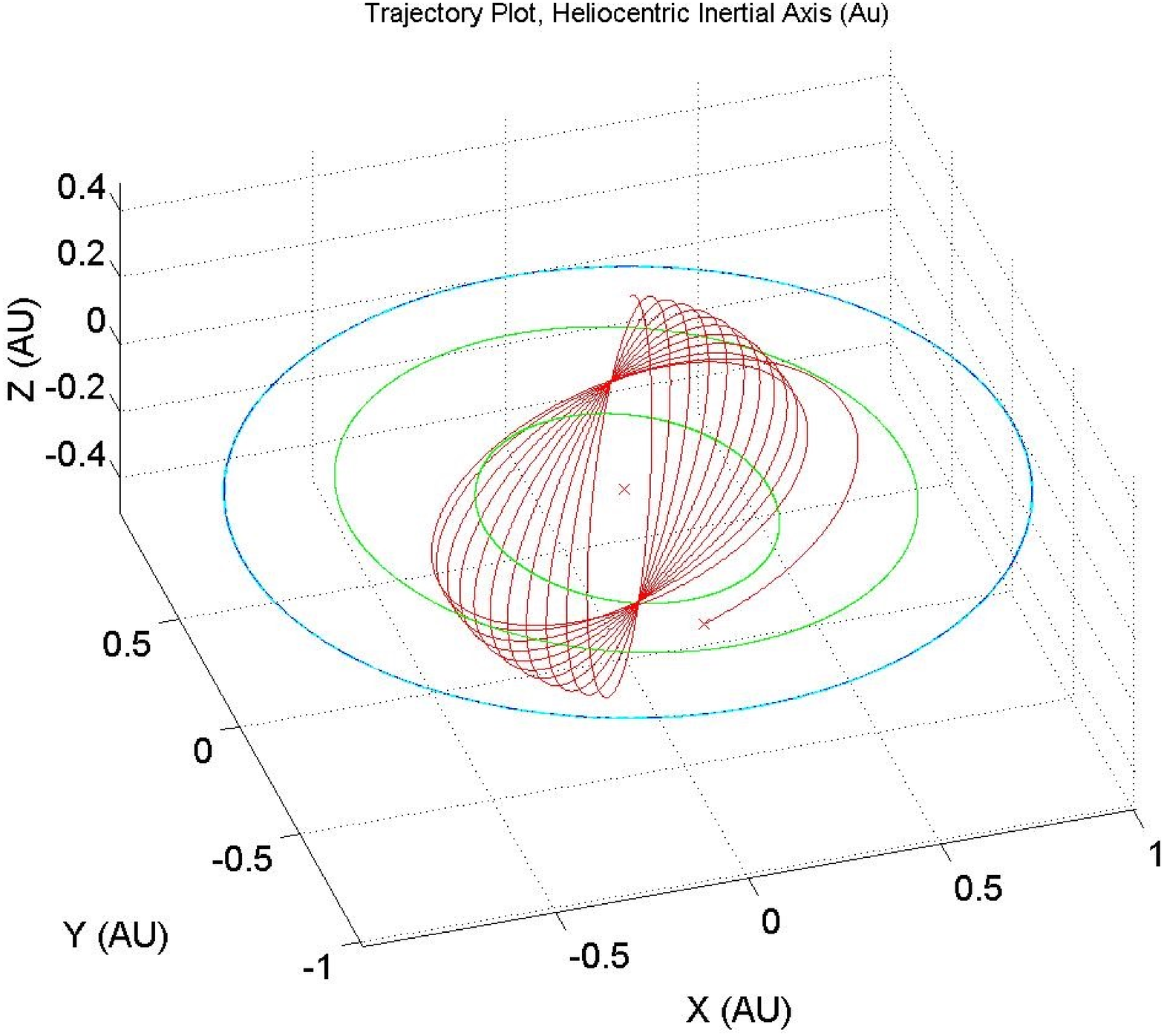}
\caption{Post-Venus Gravity Assist solar sail trajectory to a polar orbit. The orbits of Earth, Venus, and Mercury are shown.}
\label{fig:7}       
\end{figure}

 An option for a polar mission is to exploit future development of solar-sail technology to achieve a far more favourable orbital scenario much closer to the Sun and orbiting at high helio-latitudes for long periods. The concept proposed for the POLARIS mission \cite{polaris} uses a combination of a Venus Gravity Assist (VGA) and solar sail propulsion to place a spacecraft in a 0.48 AU circular orbit around the Sun with an inclination of 75$^\circ$  from the solar equator (see Figure~\ref{fig:7}). 
In this orbit, at least 59\,\% of the time would be spent at latitudes higher than the maximum latitude reached by Solar Orbiter. However, the sail size is significantly larger than has been achieved to date. So, whilst the solar sail option is considered the baseline for this concept, based on the POLARIS studies, the technological advances required are acknowledged.

\begin{figure}
  \includegraphics[scale=1.55]{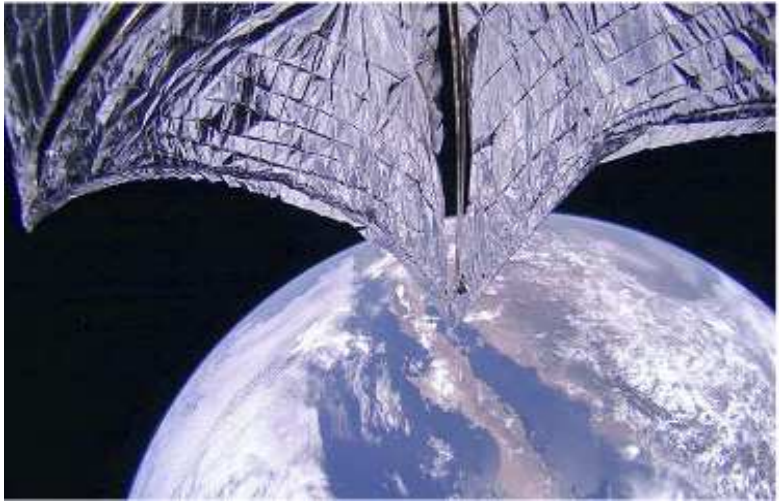}
\caption{An image taken on 25th July 2019 showing the successful deployment of the LightSail2 solar sail. (Image credit  Planetary Society)  }
\label{fig:8}       
\end{figure}

Figure~\ref{fig:8} shows the most recent solar sail mission, LightSail2 \cite{lightsail}, a crowdfunded project through the Planetary Society. The 32\,m$^2$ sail was deployed successfully in the Earth-orbiting mission to demonstrate orbital maneuvers using sail technology. The first successful solar sail technology demonstrator, IKAROS, was launched in 2010 with a 196\,m$^2$ sail. It completed its 5-year mission after an interplanetary journey that included a Venus encounter. Its principal aims were to demonstrate deployment and control of a large, thin solar sail, including attitude control through reflectance variation. Whereas the solar sail option does not require carrying a large mass of propellant, it does require storage and deployment of a solar sail over two orders of magnitude larger than any previously flown.

\subsubsection{Ion drive option}

Ion engines have been developed for many years and are increasingly used for science and exploration. A recent example is the use of four 5\,kW T6 gridded ion engines developed by QinetiQ for ESA's BepiColombo mission. ESA is continuing the study of future technologies that would be necessary for exploration missions.  These include studies on using Hall effect thrusters to keep a station around the Moon (CISLUNAR) \cite{cislunar}. Technologies like this could be explored to provide the long duration needs over the poles to obtain the science goals. 

\subsection{The way forward for technology}

In the case that the solar sail option is selected, a potential small solar sail mission technology demonstrator could be flown as a precursor to the main mission. This would be based upon a light platform and a solar sail system of reasonable size ($>$10$\times$10\,m$^2$) for a small launcher (Vega like). This mission could carry a minimal payload that would bring significant scientific return given the technology development. The precursor mission would qualify in-flight all the required technology for the solar sail system. Possibilities for instruments for this option that have $\approx$\,2\,kg mass are a small EUV imager, a magnetometer, or a total solar irradiance monitor.

Although we are talking about instrument payload options that look very much like heritage instruments from e.g. SOHO, STEREO, and Solar Orbiter, there would be a need for miniaturization of instruments given the difficulties of this orbit whilst maintaining scientific capability, and such miniaturisation would be required for all of the orbital scenario options. There would be trade-offs, e.g. between payload mass and the sail size required, for the solar sail option. 

\begin{figure}
  \includegraphics[scale=.35]{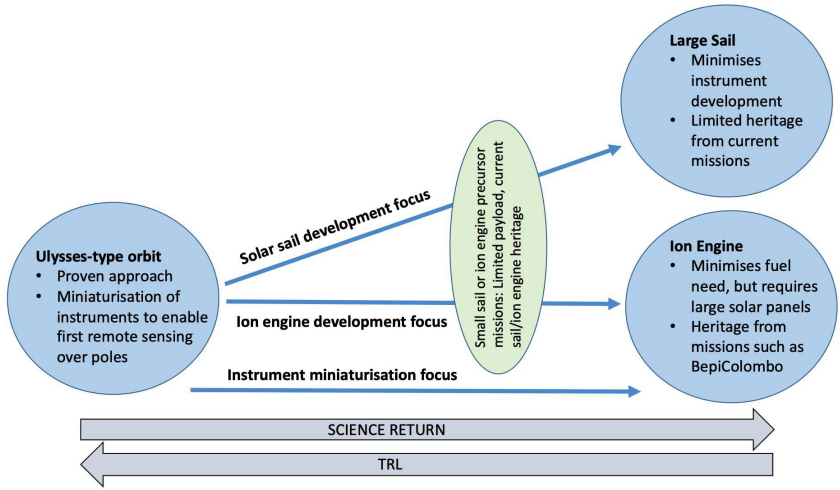}
\caption{A schematic view of the mission scenarios described and strategy.  }
\label{fig:9}       
\end{figure}

Figure~\ref{fig:9} summarises the options and challenges listed above. Since this article considers missions for the period 2035\,--\,2050 we expect that technology for sails, ion engines, and miniaturization will have progressed enough for serious consideration. The abscissa of Figure~\ref{fig:9} relates to increasing TRL to the left and increasing scientific return to the right. The Ulysses-style option is based on a proven approach, i.e. the highest TRL, and, whilst its scientific return would be extremely significant, if not ground-breaking (because we are taking remote sensing observations from high helio-latitudes for the first time), the long-duration polar observation periods and solar vicinity afforded by the sail and ion drive options would likely provide much greater scientific return. Thus, those options are shown to the right, with arrows indicating the technology development strands required. The lighter orange oval shows an opportunity for a precursor mission as the technologies are developed. For all options we are looking for instrument miniaturisation to varying degrees.

Thus, solar sails, ion engines and miniaturization of instrumentation are key to reaching the challenges of long-duration, $<$ 1 AU, high helio-latitude solar orbits. For some years there has been a drive to smaller, lower mass instruments; consider, for example, the payloads of ESA's SOHO and Solar Orbiter, which contain some similar instrument-types but with very different instrument parameters due to the use of different materials and modified optical designs. We would anticipate that this effort would continue within the instrument community.

The solar sail option would be a major challenge for the platform. Key developments are required in the following areas of solar sail technology: material, deployment (booms), attitude and orbit control system, and jettison mechanism. Some of these are discussed in \cite{Macdonald}. Different orbital concepts are presented by \cite{polaris,liewer2008}. Both employ a solar sail to reach a near-polar heliocentric orbit. One study includes a VGA to enhance transfer performance and allow for increased mass budget. A mission with a full payload using only a solar sail, requires a sail area of up to 200\,m$^2$. This would need significant technology development but, over the timescale extending to the end of the Voyage 2050 programme, we feel this should be an area of significant development in any case, as it would open up opportunities beyond solar and heliospheric physics.

With regard to ion engines, such an option was explored in the early phase of the Solar Orbiter mission, but in the end was not used, and it was considered for the POLARIS study but rejected in favour of sails. Since this time, significant developments have taken place; for example, the BepiColombo spacecraft launched in 2018 with a state-of-the-art ion engine. However this option was not deemed viable for Solar Orbiter due to the mass impact. Thus, in the context of this White Paper, an ion engine should be explored early on as an option for the mission. A ``Tundra'' type orbit could increase the fraction of the orbit at high helio-latitude.

\section{Summary and worldwide context}

There is a clear scientific drive for the international community to study the solar poles in a consistent way, over an entire activity cycle.  This is a critical next step in the global solar physics space mission program. Prior to 2006, all solar remote sensing observations were made from the ecliptic plane, on or near Earth (Earth orbit or L1). STEREO (2006 to date) provided a step change by pioneering observations out of the Sun--Earth line, and Parker Solar Probe (2018 to date) and Solar Orbiter (2020 to date) have taken the next key step by targeting the first solar encounter observations. The remaining unexplored regions are the polar regions of the Sun, targeted by this paper.  Indeed, the fact that the polar regions are not well understood is particularly unfortunate as the Earth spends a significant time magnetically connected to the polar coronal holes \cite{Luhmann}, which are difficult to observe from Earth \cite{gordon}.  
In a response to a recent international review of Next Generation Solar Physics Missions carried out by JAXA, NASA, and ESA, a range of white papers were submitted, two of which proposed polar missions \cite{NGSPM}. \cite{gibson} also recently highlighted the importance of a polar view to answer fundamental questions of the kind addressed here. The next step for a polar mission will be building on the foundation laid by Solar Orbiter, which will get the first view of the solar poles by reaching over 30$^\circ$  out of the ecliptic towards the end of its mission. Whilst, over the next decade, significant progress will be made, there is no doubt that these sneak previews of the polar regions will lay the foundation that leads to a fully fledged  polar mission.

%
%

\bibliographystyle{spmpsci}      
\bibliography{references}   

\begin{thebibliography}{10}
\providecommand{\url}[1]{{#1}}
\providecommand{\urlprefix}{URL }
\expandafter\ifx\csname urlstyle\endcsname\relax
  \providecommand{\doi}[1]{DOI~\discretionary{}{}{}#1}\else
  \providecommand{\doi}{DOI~\discretionary{}{}{}\begingroup
  \urlstyle{rm}\Url}\fi

\bibitem{aarnio2011}
{Aarnio}, A.N., {Stassun}, K.G., {Hughes}, W.J., {McGregor}, S.L.: {Solar
  Flares and Coronal Mass Ejections: A Statistically Determined Flare Flux -
  CME Mass Correlation}.
\newblock \solphys \textbf{268}(1), 195--212 (2011).
\newblock \doi{10.1007/s11207-010-9672-7}

\bibitem{polaris}
{Appourchaux}, T., {Liewer}, P., {Watt}, M., {Alexander}, D., {Andretta}, V.,
  {Auch{\`e}re}, F., {D'Arrigo}, P., {Ayon}, J., {Corbard}, T., {Fineschi}, S.,
  {Finsterle}, W., {Floyd}, L., {Garbe}, G., {Gizon}, L., {Hassler}, D.,
  {Harra}, L., {Kosovichev}, A., {Leibacher}, J., {Leipold}, M., {Murphy}, N.,
  {Maksimovic}, M., {Martinez-Pillet}, V., {Matthews}, B.S.A., {Mewaldt}, R.,
  {Moses}, D., {Newmark}, J., {R{\'e}gnier}, S., {Schmutz}, W., {Socker}, D.,
  {Spadaro}, D., {Stuttard}, M., {Trosseille}, C., {Ulrich}, R., {Velli}, M.,
  {Vourlidas}, A., {Wimmer-Schweingruber}, C.R., {Zurbuchen}, T.: {POLAR
  investigation of the Sun{\textemdash}POLARIS}.
\newblock Experimental Astronomy \textbf{23}(3), 1079--1117 (2009).
\newblock \doi{10.1007/s10686-008-9107-8}

\bibitem{Benomar}
{Benomar}, O., {Bazot}, M., {Nielsen}, M.B., {Gizon}, L., {Sekii}, T.,
  {Takata}, M., {Hotta}, H., {Hanasoge}, S., {Sreenivasan}, K.R.,
  {Christensen-Dalsgaard}, J.: {Asteroseismic detection of latitudinal
  differential rotation in 13 Sun-like stars}.
\newblock Science \textbf{361}(6408), 1231--1234 (2018).
\newblock \doi{10.1126/science.aao6571}

\bibitem{bothmer1998}
{Bothmer}, V., {Bougeret}, J.L., {Cargill}, P., {Davila}, J., {Delaboudiniere},
  J.P., {Harrison}, R., {Koutchmy}, S., {Liewer}, P., {Maltby}, P., {Rust}, D.,
  {Schwenn}, R.: {European Plans for the Solar/Heliospheric Stereo Mission}.
\newblock In: Crossroads for European Solar and Heliospheric Physics. Recent
  Achievements and Future Mission Possibilities, \emph{ESA Special
  Publication}, vol. 417, p. 145 (1998)

\bibitem{brun2017}
{Brun}, A.S., {Browning}, M.K.: {Magnetism, dynamo action and the solar-stellar
  connection}.
\newblock Liv. Rev. Solar Phys. \textbf{14}(1), 4 (2017).
\newblock \doi{10.1007/s41116-017-0007-8}

\bibitem{brun2015}
{Brun}, A.S., {Garc{\'\i}a}, R.A., {Houdek}, G., {Nandy}, D., {Pinsonneault},
  M.: {The Solar-Stellar Connection}.
\newblock \ssr \textbf{196}(1-4), 303--356 (2015).
\newblock \doi{10.1007/s11214-014-0117-8}

\bibitem{cameron}
{Cameron}, R.H., {Dikpati}, M., {Brandenburg}, A.: {The Global Solar Dynamo}.
\newblock \ssr \textbf{210}(1-4), 367--395 (2017).
\newblock \doi{10.1007/s11214-015-0230-3}

\bibitem{charbonneau}
{Charbonneau}, P.: {Dynamo Models of the Solar Cycle}.
\newblock Liv. Rev. Solar Phys. \textbf{7}(1), 3 (2010).
\newblock \doi{10.12942/lrsp-2010-3}

\bibitem{choudhuri2021}
{Choudhuri}, A.R.: {The meridional circulation of the Sun: Observations, theory
  and connections with the solar dynamo}.
\newblock Science China Physics, Mechanics, and Astronomy \textbf{64}(3),
  239601 (2021).
\newblock \doi{10.1007/s11433-020-1628-1}

\bibitem{jcd2001}
{{Christensen-Dalsgaard}, J.}: {Helioseismology: Theory}.
\newblock {Encyclopedia of astronomy and astrophysics} p. 2249 (2001)

\bibitem{corbard}
{Corbard}, T., {Thompson}, M.J.: {The subsurface radial gradient of solar
  angular velocity from MDI f-mode observations}.
\newblock \solphys \textbf{205}(2), 211--229 (2002).
\newblock \doi{10.1023/A:1014224523374}

\bibitem{Cranmer}
{Cranmer}, S.R., {van Ballegooijen}, A.A.: {On the Generation, Propagation, and
  Reflection of Alfv{\'e}n Waves from the Solar Photosphere to the Distant
  Heliosphere}.
\newblock \apjs \textbf{156}(2), 265--293 (2005).
\newblock \doi{10.1086/426507}

\bibitem{SOHO}
{Domingo}, V., {Fleck}, B., {Poland}, A.I.: {The SOHO Mission: an Overview}.
\newblock \solphys \textbf{162}(1-2), 1--37 (1995).
\newblock \doi{10.1007/BF00733425}

\bibitem{SEP}
{Dresing}, N., {G{\'o}mez-Herrero}, R., {Klassen}, A., {Heber}, B.,
  {Kartavykh}, Y., {Dr{\"o}ge}, W.: {The Large Longitudinal Spread of Solar
  Energetic Particles During the 17 January 2010 Solar Event}.
\newblock \solphys \textbf{281}(1), 281--300 (2012).
\newblock \doi{10.1007/s11207-012-0049-y}

\bibitem{Ermolli}
{Ermolli}, I., {Matthes}, K., {Dudok de Wit}, T., {Krivova}, N.A., {Tourpali},
  K., {Weber}, M., {Unruh}, Y.C., {Gray}, L., {Langematz}, U., {Pilewskie}, P.,
  {Rozanov}, E., {Schmutz}, W., {Shapiro}, A., {Solanki}, S.K., {Woods}, T.N.:
  {Recent variability of the solar spectral irradiance and its impact on
  climate modelling}.
\newblock Atmos. Chem. Phys. \textbf{13}(8), 3945--3977 (2013).
\newblock \doi{10.5194/acp-13-3945-2013}

\bibitem{finley}
{Finley}, A.J., {Matt}, S.P., {See}, V.: {The Effect of Magnetic Variability on
  Stellar Angular Momentum Loss. I. The Solar Wind Torque during Sunspot Cycles
  23 and 24}.
\newblock \apj \textbf{864}(2), 125 (2018).
\newblock \doi{10.3847/1538-4357/aad7b6}

\bibitem{fontenla1999}
{Fontenla}, J., {White}, O.R., {Fox}, P.A., {Avrett}, E.H., {Kurucz}, R.L.:
  {Calculation of Solar Irradiances. I. Synthesis of the Solar Spectrum}.
\newblock \apj \textbf{518}(1), 480--499 (1999).
\newblock \doi{10.1086/307258}

\bibitem{fontenla2009}
{Fontenla}, J.M., {Curdt}, W., {Haberreiter}, M., {Harder}, J., {Tian}, H.:
  {Semiempirical Models of the Solar Atmosphere. III. Set of Non-LTE Models for
  Far-Ultraviolet/Extreme-Ultraviolet Irradiance Computation}.
\newblock \apj \textbf{707}(1), 482--502 (2009).
\newblock \doi{10.1088/0004-637X/707/1/482}

\bibitem{Fox}
{Fox}, N.J., {Velli}, M.C., {Bale}, S.D., {Decker}, R., {Driesman}, A.,
  {Howard}, R.A., {Kasper}, J.C., {Kinnison}, J., {Kusterer}, M., {Lario}, D.,
  {Lockwood}, M.K., {McComas}, D.J., {Raouafi}, N.E., {Szabo}, A.: {The Solar
  Probe Plus Mission: Humanity's First Visit to Our Star}.
\newblock \ssr \textbf{204}(1-4), 7--48 (2016).
\newblock \doi{10.1007/s11214-015-0211-6}

\bibitem{garraffo}
{Garraffo}, C., {Drake}, J.J., {Cohen}, O.: {Magnetic Complexity as an
  Explanation for Bimodal Rotation Populations among Young Stars}.
\newblock \apjl \textbf{807}(1), L6 (2015).
\newblock \doi{10.1088/2041-8205/807/1/L6}

\bibitem{gibson}
{Gibson}, S.E., {Vourlidas}, A., {Hassler}, D.M., {Rachmeler}, L.A.,
  {Thompson}, M.J., {Newmark}, J., {Velli}, M., {Title}, A., {McIntosh}, S.W.:
  {Solar Physics from Unconventional Viewpoints}.
\newblock Front. Astron. Space Sci. \textbf{5}, 32 (2018).
\newblock \doi{10.3389/fspas.2018.00032}

\bibitem{gizon}
{Gizon}, L., {Birch}, A.C., {Spruit}, H.C.: {Local Helioseismology:
  Three-Dimensional Imaging of the Solar Interior}.
\newblock \araa \textbf{48}, 289--338 (2010).
\newblock \doi{10.1146/annurev-astro-082708-101722}

\bibitem{Gizon2020}
Gizon, L., Cameron, R.H., Pourabdian, M., Liang, Z.C., Fournier, D., Birch,
  A.C., Hanson, C.S.: {Meridional flow in the Sun’s convection zone is a
  single cell in each hemisphere}.
\newblock Science \textbf{368}(6498), 1469--1472 (2020).
\newblock \doi{10.1126/science.aaz7119}

\bibitem{margit}
{Haberreiter}, M., {Delouille}, V., {Mampaey}, B., {Verbeeck}, C., {Del Zanna},
  G., {Wieman}, S.: {Reconstruction of the solar EUV irradiance from 1996 to
  2010 based on SOHO/EIT images}.
\newblock J. Space Weather Space Clim. \textbf{4}, A30 (2014).
\newblock \doi{10.1051/swsc/2014027}

\bibitem{Hathaway2020}
{Hathaway}, D.H., {Upton}, L.A.: {Hydrodynamic Properties of the Sun's Giant
  Cellular Flows}.
\newblock arXiv e-prints arXiv:2006.06084 (2020)

\bibitem{stereo}
{Howard}, R.A., {Moses}, J.D., {Vourlidas}, A., {Newmark}, J.S., {Socker},
  D.G., {Plunkett}, S.P., {Korendyke}, C.M., {Cook}, J.W., {Hurley}, A.,
  {Davila}, J.M., {Thompson}, W.T., {St Cyr}, O.C., {Mentzell}, E., {Mehalick},
  K., {Lemen}, J.R., {Wuelser}, J.P., {Duncan}, D.W., {Tarbell}, T.D.,
  {Wolfson}, C.J., {Moore}, A., {Harrison}, R.A., {Waltham}, N.R., {Lang}, J.,
  {Davis}, C.J., {Eyles}, C.J., {Mapson-Menard}, H., {Simnett}, G.M., {Halain},
  J.P., {Defise}, J.M., {Mazy}, E., {Rochus}, P., {Mercier}, R., {Ravet}, M.F.,
  {Delmotte}, F., {Auchere}, F., {Delaboudiniere}, J.P., {Bothmer}, V.,
  {Deutsch}, W., {Wang}, D., {Rich}, N., {Cooper}, S., {Stephens}, V., {Maahs},
  G., {Baugh}, R., {McMullin}, D., {Carter}, T.: {Sun Earth Connection Coronal
  and Heliospheric Investigation (SECCHI)}.
\newblock \ssr \textbf{136}(1-4), 67--115 (2008).
\newblock \doi{10.1007/s11214-008-9341-4}

\bibitem{howelrsp}
{Howe}, R.: {Solar Interior Rotation and its Variation}.
\newblock Living Reviews in Solar Physics \textbf{6}(1), 1 (2009).
\newblock \doi{10.12942/lrsp-2009-1}

\bibitem{kilpua}
{Kilpua}, E.K.J., {Balogh}, A., {von Steiger}, R., {Liu}, Y.D.: {Geoeffective
  Properties of Solar Transients and Stream Interaction Regions}.
\newblock \ssr \textbf{212}(3-4), 1271--1314 (2017).
\newblock \doi{10.1007/s11214-017-0411-3}

\bibitem{Knaack}
{Knaack}, R., {Fligge}, M., {Solanki}, S.K., {Unruh}, Y.C.: {The influence of
  an inclined rotation axis on solar irradiance variations}.
\newblock \aap \textbf{376}, 1080--1089 (2001).
\newblock \doi{10.1051/0004-6361:20011007}

\bibitem{koomen1974}
{Koomen}, M., {Howard}, R., {Hansen}, R., {Hansen}, S.: {The Coronal Transient
  of 16 June 1972}.
\newblock \solphys \textbf{34}(2), 447--452 (1974).
\newblock \doi{10.1007/BF00153680}

\bibitem{Hinode}
{Kosugi}, T., {Matsuzaki}, K., {Sakao}, T., {Shimizu}, T., {Sone}, Y.,
  {Tachikawa}, S., {Hashimoto}, T., {Minesugi}, K., {Ohnishi}, A., {Yamada},
  T., {Tsuneta}, S., {Hara}, H., {Ichimoto}, K., {Suematsu}, Y., {Shimojo}, M.,
  {Watanabe}, T., {Shimada}, S., {Davis}, J.M., {Hill}, L.D., {Owens}, J.K.,
  {Title}, A.M., {Culhane}, J.L., {Harra}, L.K., {Doschek}, G.A., {Golub}, L.:
  {The Hinode (Solar-B) Mission: An Overview}.
\newblock \solphys \textbf{243}(1), 3--17 (2007).
\newblock \doi{10.1007/s11207-007-9014-6}

\bibitem{kraft}
{Kraft}, R.P.: {Studies of Stellar Rotation. V. The Dependence of Rotation on
  Age among Solar-Type Stars}.
\newblock \apj \textbf{150}, 551 (1967).
\newblock \doi{10.1086/149359}

\bibitem{kwon}
{Kwon}, R.Y., {Vourlidas}, A.: {The density compression ratio of shock fronts
  associated with coronal mass ejections}.
\newblock J. Space Weather Space Clim. \textbf{8}, A08 (2018).
\newblock \doi{10.1051/swsc/2017045}

\bibitem{lanzafame}
{Lanzafame}, A.C., {Spada}, F.: {Rotational evolution of slow-rotator sequence
  stars}.
\newblock \aap \textbf{584}, A30 (2015).
\newblock \doi{10.1051/0004-6361/201526770}

\bibitem{ramses1998}
{Le Qu{\'e}au}, D., {Roux}, A., {Vial}, J.C.: {Solar Probe: The RAMSES
  Proposal}.
\newblock In: Crossroads for European Solar and Heliospheric Physics. Recent
  Achievements and Future Mission Possibilities, \emph{ESA Spec. Pub.}, vol.
  417, p.~75 (1998)

\bibitem{liang2019}
{Liang}, Z.C., {Gizon}, L., {Birch}, A.C., {Duvall}, T.L.: {Time-distance
  helioseismology of solar Rossby waves}.
\newblock \aap \textbf{626}, A3 (2019).
\newblock \doi{10.1051/0004-6361/201834849}

\bibitem{liewer2008}
{Liewer}, P.C., {Ayon}, J., {Alexander}, D., {Kosovichev}, A., {Mewaldt}, R.A.,
  {Socker}, D.G., {Vourlidas}, A.: {Solar Polar Imager: Observing Solar
  Activity from a New Perspective}, vol. 224, p.~1.
\newblock Nat. Res. Council (2008)

\bibitem{lightsail}
{LightSail Team}: Website of {L}ight{S}ail.
\newblock \url{https://www.planetary.org/sci-tech/lightsail} (2021).
\newblock {Online; }

\bibitem{Linker}
{Linker}, J.A., {Caplan}, R.M., {Downs}, C., {Riley}, P., {Mikic}, Z.,
  {Lionello}, R., {Henney}, C.J., {Arge}, C.N., {Liu}, Y., {Derosa}, M.L.,
  {Yeates}, A., {Owens}, M.J.: {The Open Flux Problem}.
\newblock \apj \textbf{848}(1), 70 (2017).
\newblock \doi{10.3847/1538-4357/aa8a70}

\bibitem{liu}
{Liu}, Y.D., {Luhmann}, J.G., {Kajdi{\v{c}}}, P., {Kilpua}, E.K.J., {Lugaz},
  N., {Nitta}, N.V., {M{\"o}stl}, C., {Lavraud}, B., {Bale}, S.D., {Farrugia},
  C.J., {Galvin}, A.B.: {Observations of an extreme storm in interplanetary
  space caused by successive coronal mass ejections}.
\newblock Nature Comm. \textbf{5}, 3481 (2014).
\newblock \doi{10.1038/ncomms4481}

\bibitem{Lockwood}
{Lockwood}, G.W., {Skiff}, B.A., {Baliunas}, S.L., {Radick}, R.R.: {Long-term
  solar brightness changes estimated from a survey of Sun-like stars}.
\newblock \nat \textbf{360}(6405), 653--655 (1992).
\newblock \doi{10.1038/360653a0}

\bibitem{loeptien2018}
{L{\"o}ptien}, B., {Gizon}, L., {Birch}, A.C., {Schou}, J., {Proxauf}, B.,
  {Duvall}, T.L., {Bogart}, R.S., {Christensen}, U.R.: {Global-scale equatorial
  Rossby waves as an essential component of solar internal dynamics}.
\newblock Nature Astronomy \textbf{2}, 568--573 (2018).
\newblock \doi{10.1038/s41550-018-0460-x}

\bibitem{lugaz}
{Lugaz}, N., {Temmer}, M., {Wang}, Y., {Farrugia}, C.J.: {The Interaction of
  Successive Coronal Mass Ejections: A Review}.
\newblock \solphys \textbf{292}(4), 64 (2017).
\newblock \doi{10.1007/s11207-017-1091-6}

\bibitem{Luhmann}
{Luhmann}, J.G., {Lee}, C.O., {Li}, Y., {Arge}, C.N., {Galvin}, A.B.,
  {Simunac}, K., {Russell}, C.T., {Howard}, R.A., {Petrie}, G.: {Solar Wind
  Sources in the Late Declining Phase of Cycle 23: Effects of the Weak Solar
  Polar Field on High Speed Streams}.
\newblock \solphys \textbf{256}(1-2), 285--305 (2009).
\newblock \doi{10.1007/s11207-009-9354-5}

\bibitem{Macdonald}
{MacDonald}, M., {Atzei}, A., {Falkner}, P., {Hughes}, G., {Lyngvi}, A.,
  {McInnes}, C.: {Solar Polar Orbiter: A Solar Sail Technology Reference
  Study}.
\newblock J. Spacecraft Rockets \textbf{43}(5), 960--972 (2006).
\newblock \doi{10.2514/1.16408}

\bibitem{cislunar}
{Mammarella}, M., {Vernicari}, P.M., {Paissoni}, C.A., {Viola}, N.: {How the
  Lunar Space Tug can support the cislunar station}.
\newblock Acta Astronautica \textbf{154}, 181--194 (2019).
\newblock \doi{10.1016/j.actaastro.2018.04.032}

\bibitem{manchester}
{Manchester}, W., {Kilpua}, E.K.J., {Liu}, Y.D., {Lugaz}, N., {Riley}, P.,
  {T{\"o}r{\"o}k}, T., {Vr{\v{s}}nak}, B.: {The Physical Processes of CME/ICME
  Evolution}.
\newblock \ssr \textbf{212}(3-4), 1159--1219 (2017).
\newblock \doi{10.1007/s11214-017-0394-0}

\bibitem{matt}
{Matt}, S.P., {Brun}, A.S., {Baraffe}, I., {Bouvier}, J., {Chabrier}, G.: {The
  Mass-dependence of Angular Momentum Evolution in Sun-like Stars}.
\newblock \apjl \textbf{799}(2), L23 (2015).
\newblock \doi{10.1088/2041-8205/799/2/L23}

\bibitem{McComas1998}
{McComas}, D.J., {Bame}, S.J., {Barraclough}, B.L., {Feldman}, W.C., {Funsten},
  H.O., {Gosling}, J.T., {Riley}, P., {Skoug}, R., {Balogh}, A., {Forsyth}, R.,
  {Goldstein}, B.E., {Neugebauer}, M.: {Ulysses' return to the slow solar
  wind}.
\newblock \grl \textbf{25}(1), 1--4 (1998).
\newblock \doi{10.1029/97GL03444}

\bibitem{mccomas2003}
{McComas}, D.J., {Elliott}, H.A., {Schwadron}, N.A., {Gosling}, J.T., {Skoug},
  R.M., {Goldstein}, B.E.: {The three-dimensional solar wind around solar
  maximum}.
\newblock \grl \textbf{30}(10), 1517 (2003).
\newblock \doi{10.1029/2003GL017136}

\bibitem{Mestel}
{Mestel}, L.: {Magnetic braking by a stellar wind-I}.
\newblock \mnras \textbf{138}, 359 (1968).
\newblock \doi{10.1093/mnras/138.3.359}

\bibitem{mishra}
{Mishra}, W., {Srivastava}, N., {Wang}, Y., {Mirtoshev}, Z., {Zhang}, J.,
  {Liu}, R.: {Mass loss via solar wind and coronal mass ejections during solar
  cycles 23 and 24}.
\newblock \mnras \textbf{486}(4), 4671--4685 (2019).
\newblock \doi{10.1093/mnras/stz1001}

\bibitem{orbiter2020}
{M{\"u}ller}, D., {St. Cyr}, O.C., {Zouganelis}, I., {Gilbert}, H.R.,
  {Marsden}, R., {Nieves-Chinchilla}, T., {Antonucci}, E., {Auch{\`e}re}, F.,
  {Berghmans}, D., {Horbury}, T.S., {Howard}, R.A., {Krucker}, S.,
  {Maksimovic}, M., {Owen}, C.J., {Rochus}, P., {Rodriguez-Pacheco}, J.,
  {Romoli}, M., {Solanki}, S.K., {Bruno}, R., {Carlsson}, M., {Fludra}, A.,
  {Harra}, L., {Hassler}, D.M., {Livi}, S., {Louarn}, P., {Peter}, H.,
  {Sch{\"u}hle}, U., {Teriaca}, L., {del Toro Iniesta}, J.C.,
  {Wimmer-Schweingruber}, R.F., {Marsch}, E., {Velli}, M., {De Groof}, A.,
  {Walsh}, A., {Williams}, D.: {The Solar Orbiter mission. Science overview}.
\newblock \aap \textbf{642}, A1 (2020).
\newblock \doi{10.1051/0004-6361/202038467}

\bibitem{NGSPM}
{{N}{G}{S}{P}{M} {S}cience {O}bjectives {T}eam}: Next generation solar physics
  mission science objectives team (ngspm-so).
\newblock
  \url{https://hinode.nao.ac.jp/SOLAR-C/SOLAR-C/Documents/NGSPM_report_170731.pdf}
  (2017).
\newblock [Online]

\bibitem{SDO}
{Pesnell}, W.D., {Thompson}, B.J., {Chamberlin}, P.C.: {The Solar Dynamics
  Observatory (SDO)}.
\newblock \solphys \textbf{275}(1-2), 3--15 (2012).
\newblock \doi{10.1007/s11207-011-9841-3}

\bibitem{gordon}
{Petrie}, G.J.D.: {Solar Magnetism in the Polar Regions}.
\newblock Liv. Rev. Solar Phys. \textbf{12}(1), 5 (2015).
\newblock \doi{10.1007/lrsp-2015-5}

\bibitem{Radick}
{Radick}, R.R., {Lockwood}, G.W., {Henry}, G.W., {Hall}, J.C., {Pevtsov}, A.A.:
  {Patterns of Variation for the Sun and Sun-like Stars}.
\newblock \apj \textbf{855}(2), 75 (2018).
\newblock \doi{10.3847/1538-4357/aaaae3}

\bibitem{rempel2007}
{Rempel}, M.: {Origin of Solar Torsional Oscillations}.
\newblock \apj \textbf{655}(1), 651--659 (2007).
\newblock \doi{10.1086/509866}

\bibitem{rempel2012}
{Rempel}, M.: {High-latitude Solar Torsional Oscillations during Phases of
  Changing Magnetic Cycle Amplitude}.
\newblock \apjl \textbf{750}(1), L8 (2012).
\newblock \doi{10.1088/2041-8205/750/1/L8}

\bibitem{Rosenbauer}
{Rosenbauer}, H., {Schwenn}, R., {Marsch}, E., {Meyer}, B., {Miggenrieder}, H.,
  {Montgomery}, M.D., {Muehlhaeuser}, K.H., {Pilipp}, W., {Voges}, W., {Zink},
  S.M.: {A survey on initial results of the HELIOS plasma experiment}.
\newblock J. Geophys. Zeits. Geophys. \textbf{42}(6), 561--580 (1977)

\bibitem{alexis}
{Rouillard}, A.P., {Davies}, J.A., {Forsyth}, R.J., {Rees}, A., {Davis}, C.J.,
  {Harrison}, R.A., {Lockwood}, M., {Bewsher}, D., {Crothers}, S.R., {Eyles},
  C.J., {Hapgood}, M., {Perry}, C.H.: {First imaging of corotating interaction
  regions using the STEREO spacecraft}.
\newblock \grl \textbf{35}(10), L10110 (2008).
\newblock \doi{10.1029/2008GL033767}

\bibitem{thompson}
{Thompson}, M.J., {Christensen-Dalsgaard}, J., {Miesch}, M.S., {Toomre}, J.:
  {The Internal Rotation of the Sun}.
\newblock \araa \textbf{41}, 599--643 (2003).
\newblock \doi{10.1146/annurev.astro.41.011802.094848}

\bibitem{tousey1973}
{Tousey}, R.: {The solar corona.}
\newblock In: Space Research Conference, vol.~2, pp. 713--730 (1973)

\bibitem{vennerstrom}
{Vennerstrom}, S., {Lefevre}, L., {Dumbovi{\'c}}, M., {Crosby}, N.,
  {Malandraki}, O., {Patsou}, I., {Clette}, F., {Veronig}, A., {Vr{\v{s}}nak},
  B., {Leer}, K., {Moretto}, T.: {Extreme Geomagnetic Storms - 1868 - 2010}.
\newblock \solphys \textbf{291}(5), 1447--1481 (2016).
\newblock \doi{10.1007/s11207-016-0897-y}

\bibitem{vourlidas2018}
{Vourlidas}, A., {Liewer}, P.C., {Velli}, M., {Webb}, D.: {Solar Polar Diamond
  Explorer (SPDEx): Understanding the Origins of Solar Activity Using a New
  Perspective}.
\newblock arXiv e-prints arXiv:1805.04172 (2018)

\bibitem{webb2017}
{Webb}, D.F., {Howard}, R.A., {St. Cyr}, O.C., {Vourlidas}, A.: {Is There a CME
  Rate Floor? CME and Magnetic Flux Values for the Last Four Solar Cycle
  Minima}.
\newblock \apj \textbf{851}(2), 142 (2017).
\newblock \doi{10.3847/1538-4357/aa9b81}

\bibitem{Ulysses}
{Wenzel}, K.P., {Marsden}, R.G., {Page}, D.E., {Smith}, E.J.: {The ULYSSES
  Mission}.
\newblock \aaps \textbf{92}, 207 (1992)

\bibitem{Wenzel}
{Wenzel}, K.P., {Marsden}, R.G., {Page}, D.E., {Smith}, E.J.: {The ULYSSES
  Mission}.
\newblock \aaps \textbf{92}, 207 (1992)

\bibitem{xiong2018}
{Xiong}, M., {Davies}, J.A., {Harrison}, R.A., {Zhou}, Y., {Feng}, X., {Xia},
  L., {Li}, B., {Liu}, Y.D., {Hayashi}, K., {Li}, H., {Yang}, L.: {Prospective
  Out-of-ecliptic White-light Imaging of Coronal Mass Ejections Traveling
  through the Corona and Heliosphere}.
\newblock \apj \textbf{852}(2), 111 (2018).
\newblock \doi{10.3847/1538-4357/aaa028}

\bibitem{xiong2017}
{Xiong}, M., {Davies}, J.A., {Li}, B., {Yang}, L., {Liu}, Y.D., {Xia}, L.,
  {Harrison}, R.A., {Keiji}, H., {Li}, H.: {Prospective Out-of-ecliptic
  White-light Imaging of Interplanetary Corotating Interaction Regions at Solar
  Maximum}.
\newblock \apj \textbf{844}(1), 76 (2017).
\newblock \doi{10.3847/1538-4357/aa7aaa}

\bibitem{sport}
{Xiong}, M., {Liu}, Y., {Liu}, H., {Li}, B., {Zheng}, J., {Zhang}, C., {Xia},
  L., {Zhang}, H., {Rao}, W., {Chen}, C., {Sun}, W., {Wu}, X., {Deng}, Y.,
  {He}, H., {Jiang}, B., {Wang}, Y., {Wang}, C., {Shen}, C., {Zhang}, H.,
  {Zhang}, S., {Yang}, X., {Sang}, P., {Wu}, J.: {Overview of the Solar Polar
  Orbit Telescopeproject for space weather mission}.
\newblock {Chin. J. Space Sci.} \textbf{36(3)}, 245--266 (2016).
\newblock \doi{10.11728/cjss2016.03.245}.
\newblock \urlprefix\url{{http://www.cjss.ac.cn/EN/10.11728/cjss2016.03.245}}

\bibitem{yashiro}
{Yashiro}, S., {Gopalswamy}, N., {Michalek}, G., {St. Cyr}, O.C., {Plunkett},
  S.P., {Rich}, N.B., {Howard}, R.A.: {A catalog of white light coronal mass
  ejections observed by the SOHO spacecraft}.
\newblock J. Geophys. Res. (Space Phys.) \textbf{109}(A7), A07105 (2004).
\newblock \doi{10.1029/2003JA010282}

\bibitem{Yeo}
{Yeo}, K.L., {Solanki}, S.K., {Norris}, C.M., {Beeck}, B., {Unruh}, Y.C.,
  {Krivova}, N.A.: {Solar Irradiance Variability is Caused by the Magnetic
  Activity on the Solar Surface}.
\newblock \prl \textbf{119}, 9.1102 (2017).
\newblock \doi{10.1103/PhysRevLett.119.091102}

\end{thebibliography}

\end{document}